\documentclass[a4paper,11pt]{article}
\pdfoutput=1 

\usepackage{jinstpub} 

\usepackage{lineno}

\title{\boldmath Cryogenic operation of silicon photomultiplier arrays}

\author[a,1]{E. Currás-Rivera,\note{Corresponding author.}}
\author[a]{F. Blanc, G. Haefeli, R. Marchevski, F. Ronchetti, O. Schneider, L. Shchutska and  G. Zunica}

\affiliation[a]{École Polytechnique Fédérale de Lausanne,\\Route de la Sorge, 1015 Écublens, Switzerland}

\emailAdd{esteban.curras.rivera@cern.ch}

\abstract{The LHCb experiment at CERN has been upgraded for the Run 3 operation of the Large Hadron Collider (LHC). A new concept of tracking detector based on Scintillating Fibres (SciFi) read out with multichannel silicon photomultipliers (SiPMs) was installed during its upgrade. One of the main challenges the SciFi tracker will face during the Run 4 operation of the LHC is the higher radiation environment due to fast neutrons, where the SiPMs are located. To cope with the increase in radiation, cryogenic cooling with liquid nitrogen is being investigated as a possible solution to mitigate the performance degradation of the SiPMs induced by radiation damage. Thus, a detailed performance study of different layouts of SiPM arrays produced by Fondazione Bruno Kessler (FBK) and Hamamatsu Photonics K.K. is being carried out. These SiPMs have been designed to operate at cryogenic temperatures. Several SiPMs have been tested in a dedicated cryogenic setup down to 100 K. Key performance parameters such as breakdown voltage, dark count rate, photon detection efficiency, gain and direct cross-talk are characterized as a function of the temperature. The main results of this study are going to be presented here.}


\proceeding{6$^{\text{th}}$  International Workshop on New Photon-Detector\\
  November 19-21 2024\\
  Harbour Centre, Vancouver (BC) Canada}

\begin{document}
\maketitle
\flushbottom

\section{Introduction}
\label{sec:intro}
The cryogenic operation of silicon photomultiplier detectors (SiPM) in high-energy physics experiments has been gaining interest in recent years and important developments have been made to improve their performance at cryogenic temperatures such as liquid argon, liquid xenon, or liquid nitrogen environments. The cryogenic operation of SiPMs brings several advantages for future particle detector experiments, but also new challenges must be addressed ~\cite{a,b}.

Currently, this approach is being investigated to upgrade the Scintillating Fibers (SciFi) tracker detector of the LHCb experiment at CERN ~\cite{c}. For the Run 3 operation, the SiPMs are operated at -40\,$^\circ$C to lower the radiation-induced dark count rate but, given the future upgrade of the LHCb experiment for the High-Luninosity LHC (upgrade II), this will not be enough due to the high radiation environment expected during its operation ~\cite{d}. Thus, cryogenic cooling with liquid nitrogen is being investigated, and a detailed study of SiPM performance under cryogenic operation needs to be assessed.

To perform this study a dedicated cryogenic setup was developed. This setup allows the evaluation of all the critical performance parameters against temperature from room temperature (RT) down to 100 K. Different SiPM technologies, from two different producers, that differ in the gain, pixel size, breakdown voltage, gain layer implantation parameters, and photon detection efficiency were characterized and the main results are reported here.

\section{SiPM arrays under study}
\label{sec:sipm}

The SiPM arrays under study were produced by Hamamatsu Photonics K.K. and by Fondazione Bruno Kessler (FBK). A SiPM array comprises 128 channels to cover a total area of \mbox{32.54 mm $\times$ 1.625 mm}. Each array is built using two silicon dice of 64 channels mounted on a Kapton Printed Circuit Board (PCB). Each channel has a dimension of \mbox{1.665 mm $\times$ 0.25 mm} and it is built with a total of 104 pixels for the Hamamatsu SiPMs (with a pixel size of \mbox{$62.5\times57.5\,\mu m^2$}) and 234 pixels for the FBK  SiPMs (with a pixel size of \mbox{$41.7\times41.7\,\mu m^2 $}). A quenching resistor ($R_q$) of \mbox{$500\, k \Omega $} at RT is implemented in all designs. The FBK SiPM arrays are part of the near-ultraviolet high-density (NUV-HD) generation ~\cite{e} and the Hamamatsu SiPM arrays belong to their production ID S13552 ~\cite{f}.

\section{Cryogenic set-up}
\label{sec:setup}

The measurements were performed using a temperature controller closed-cycle He cryostat system. The SiPM arrays are placed inside the cryostat chamber allowing measurements at any temperature between 100 K and 300 K (RT). To allow the control of the temperature inside the cryostat chamber, two \mbox{50 W} heaters are used. A pump system creates a vacuum inside the cryostat chamber where the SiPM arrays are located. Several feed-throughs allow access to the chamber to monitor the temperature and to perform the light injection. The monitor of the temperature is done using PT1000 thermistors located at the proximity of each SiPM array and for the light injection, a \mbox{450 nm} wavelength pulsed laser is used. The light is diffused to have a homogeneous illumination with an intensity such that the average number of detected photons is about one. The bias voltage of the SiPM arrays is applied using a Keithley 2450 source meter that is also used to monitor the current. The signal is amplified using a 40 dB amplifier with a bandwidth of 2 GHz and digitized using a 4 GHz and 20 GS/s oscilloscope.

\section{Results}
\label{sec:results}

\subsection{Breakdown voltage}

\begin{figure}[htbp]
\centering 
\includegraphics[width=.47\textwidth]{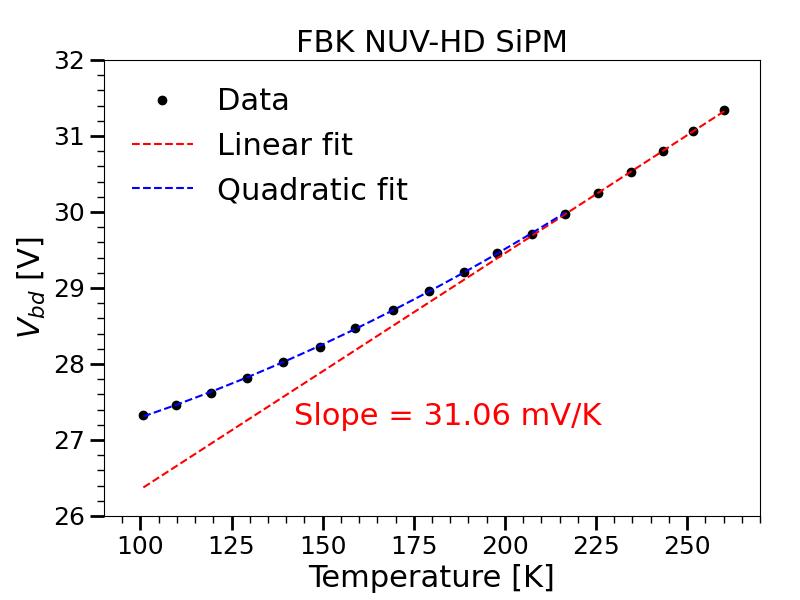}
\qquad
\includegraphics[width=.47\textwidth]{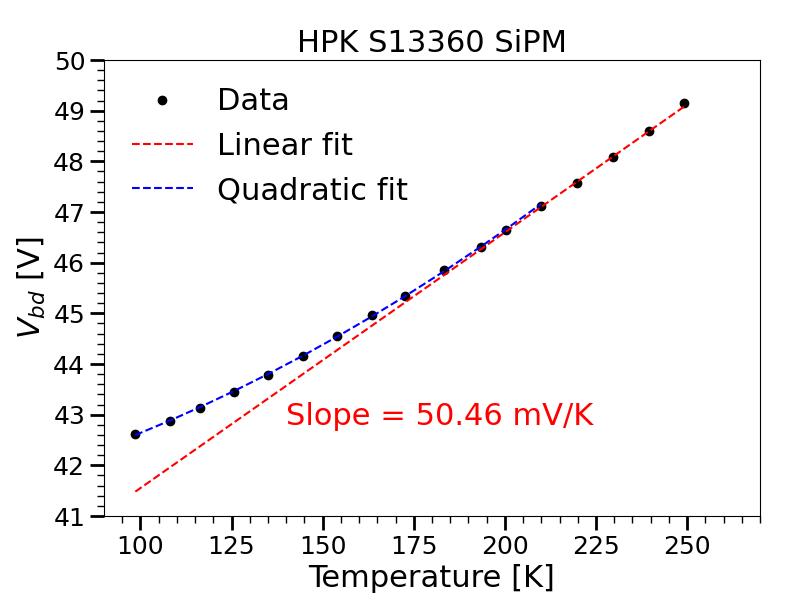}
\caption{\label{fig:vbd} Breakdown voltage against temperature for the two SiPM technologies under study.}
\end{figure}

The breakdown voltage ($V_{bd}$) was extracted from the current-voltage (I-V) characteristic using the Inverse Logarithmic Derivative (ILD) method ~\cite{g}. The I-V characteristic is measured at each temperature exposing the SiPM arrays to a constant light illumination to improve the resolution in the measurement. Furthermore, several channels from each SiPM array were measured together, in this way the extracted $V_{bd}$ represents a good approximation of the average $V_{bd}$ of the SiPM array. The obtained $V_{bd}$ against temperature for the two types of SiPM arrays under study are shown in figure~\ref{fig:vbd}.

\subsection{Quenching resistor and recovery time}

\begin{figure}[htbp]
\centering 
\includegraphics[width=.47\textwidth]{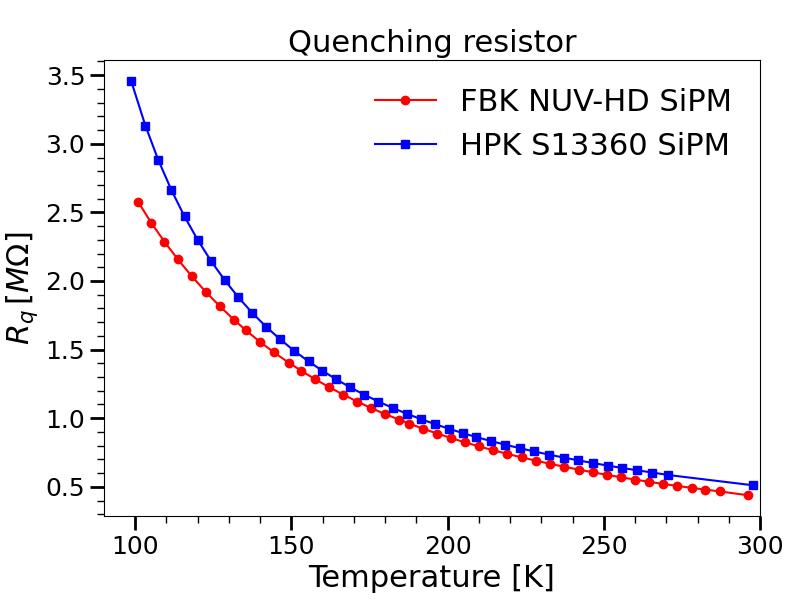}
\qquad
\includegraphics[width=.47\textwidth]{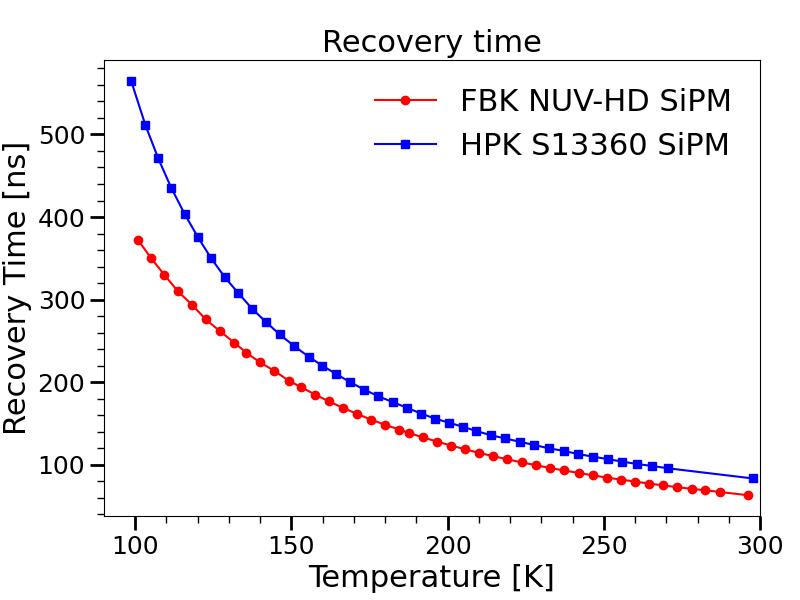}
\caption{\label{fig:rq} Quenching resistor and recovery time against temperature for the two SiPM technologies under study.}
\end{figure}

The quenching resistor value ($R_q$) is extracted from the I-V characteristics in the forward bias region, where the quenching resistor dominates over the diode characteristics and makes the response proportional to the bias voltage. This slope is used to compute the $R_q$ value. By design, all the SiPM arrays have an $R_q$ value of \mbox{$500\, K \Omega $} at RT. The obtained $R_q$ value against temperature for the two types of SiPM arrays under study are shown in figure~\ref{fig:rq}.

\subsection{Photon detection efficiency}

\begin{figure}[htbp]
\centering 
\includegraphics[width=.47\textwidth]{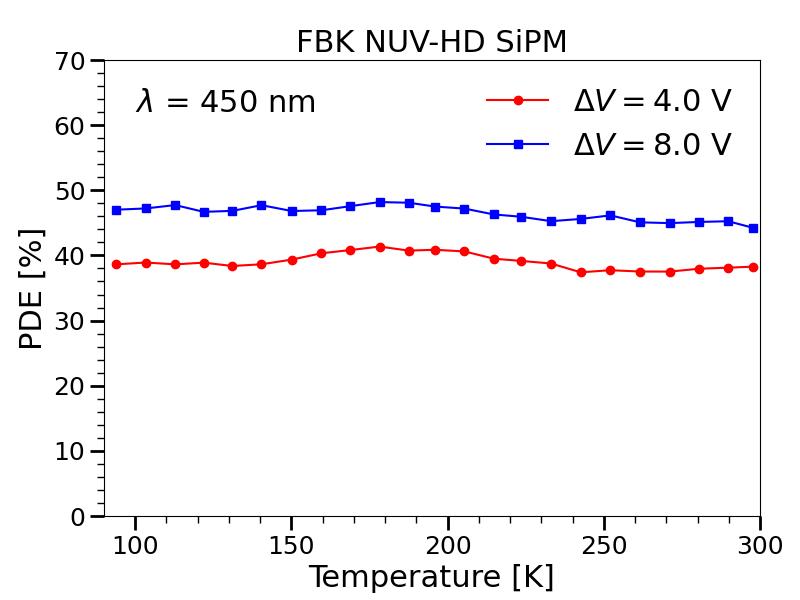}
\qquad
\includegraphics[width=.47\textwidth]{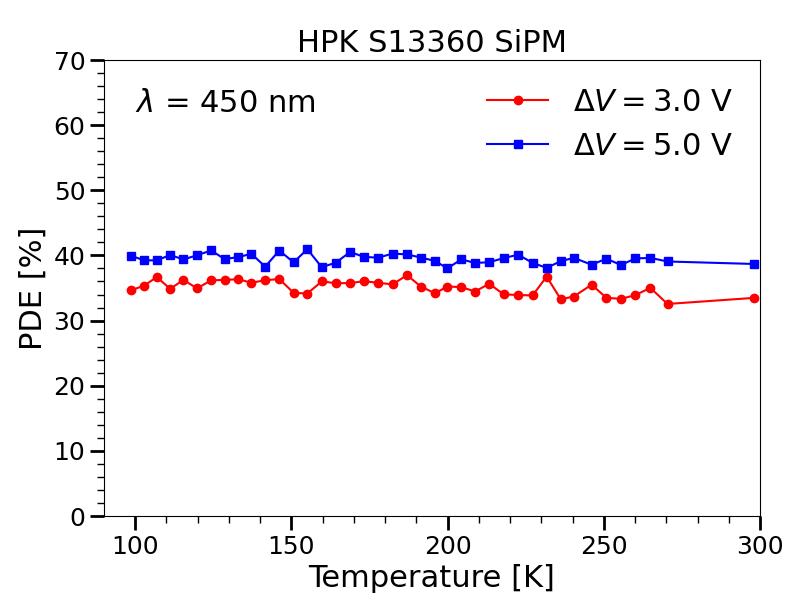}
\caption{\label{fig:pde} Photon detection efficiency against temperature for the two SiPM technologies under study.}
\end{figure}

The absolute photon detection efficiency (PDE) was first measured at RT in a dedicated setup for the two SiPM array technologies under study. More details about the setup and how the absolute PDE is obtained can be found in this reference ~\cite{h}. FBK and Hamamatsu SiPMs show different PDE spectral responses with different peak sensitivities. Nevertheless, on the cryostat, we limit the measurements to one wavelength ($\lambda$ = 450 nm). Due to the setup limitations, only relative PDE measurements can be carried out, which means that the PDE obtained is scaled to the absolute one measured at 300 K. The PDE was computed using the measured pulse rate under the laser illumination. A noise correction, mainly to subtract the contribution from after pulses (AP) and cross-talk was performed. The obtained PDE against temperature for the two types of SiPM arrays under study are shown in figure~\ref{fig:pde}. The PDE showed is limited only to two overvoltages ($\Delta V$) per SiPM array.

\subsection{Gain}

\begin{figure}[htbp]
\centering 
\includegraphics[width=.47\textwidth]{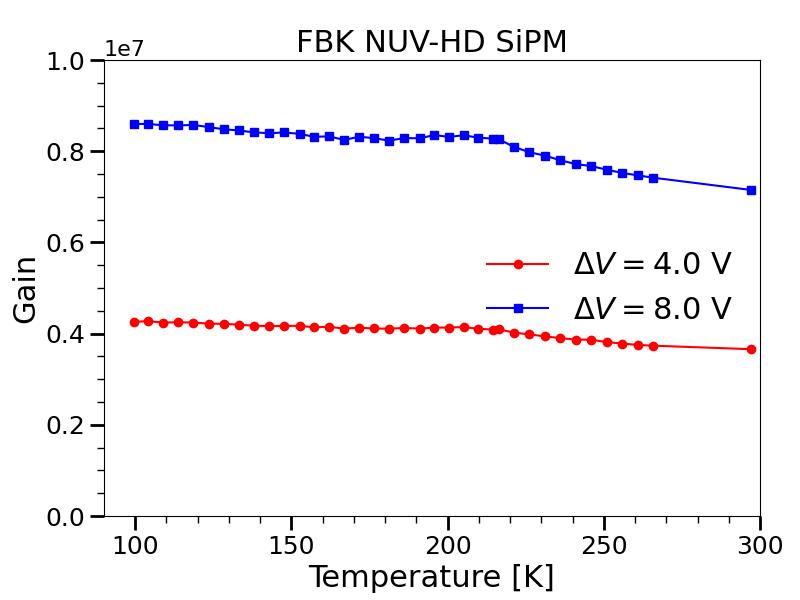}
\qquad
\includegraphics[width=.47\textwidth]{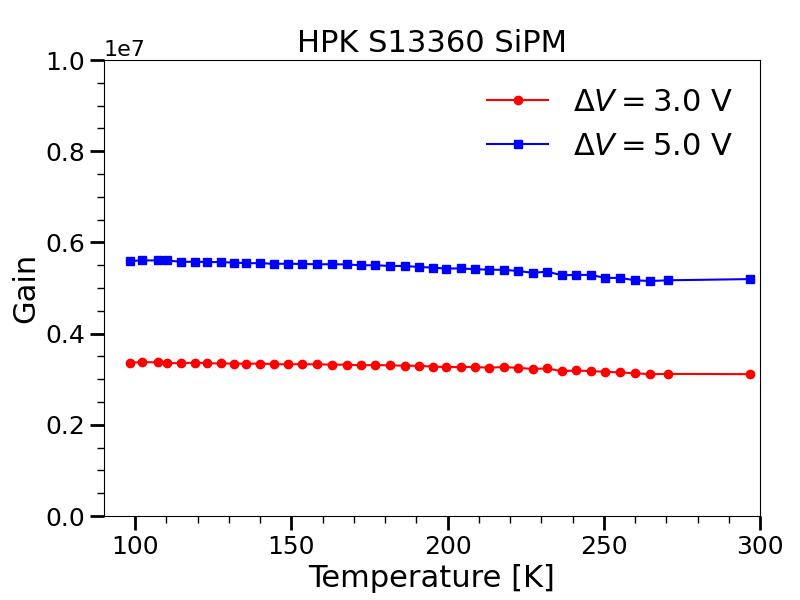}
\caption{\label{fig:fgain} Gain against temperature for the two SiPM technologies under study.}
\end{figure}

The gain was computed using the photocurrent ($I_{photo}$) and pulse rate ($R_{photo}$), measured for each $\Delta V$ and temperature (T), using equation \eqref{eq:egain}, where '$e$' represents the elementary charge. The obtained gain against temperature for the two types of SiPM arrays under study is shown in figure~\ref{fig:fgain}. The gain is shown for two different $\Delta V$, 4.0 V and 8.0 V for the FBK SiPM and, 3.0 V and 5.0 V for the HPK one. Both SiPMs show slightly different values of the gain per voltage unit: G/($\Delta V$) = $7.6\times10^{5} \pm 9.0\times10^{3}$ (FBK NUV-HD) and \mbox{$G/(\Delta V) = 1.06\times10^{6} \pm 6.0\times10^{3}$} (HPK S13360).

\begin{equation}
\label{eq:egain}
\begin{aligned}
Gain\left( \Delta V, T \right) = \frac{I_{photo}\left( \Delta V, T \right)}{R_{photo}\left( \Delta V, T \right)\times e}
\end{aligned}
\end{equation}

\subsection{Correlated noise: direct cross-talk probability}

\begin{figure}[htbp]
\centering 
\includegraphics[width=.47\textwidth]{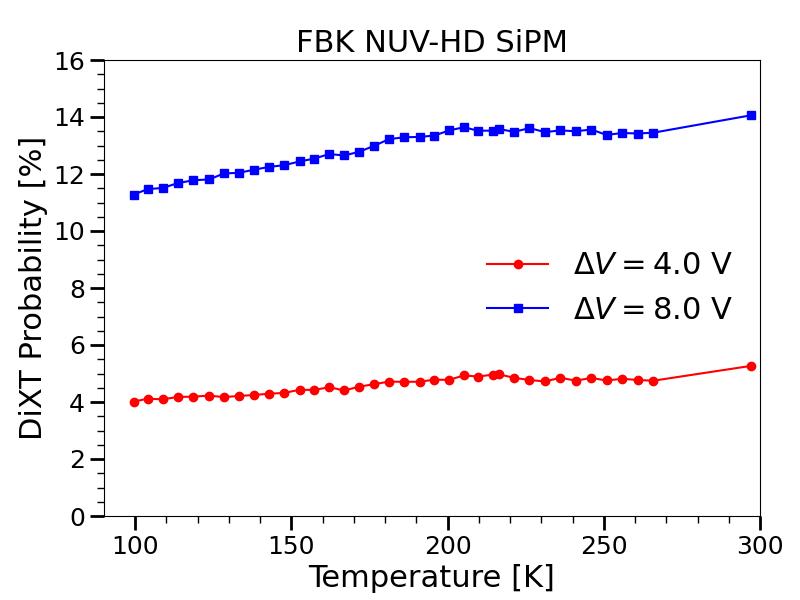}
\qquad
\includegraphics[width=.47\textwidth]{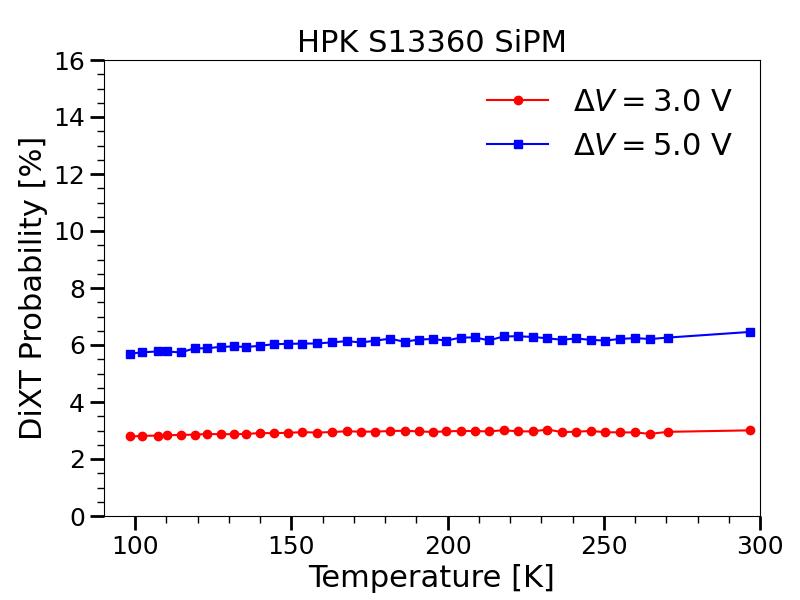}
\caption{\label{fig:dixt} Direct cross-talk probability against temperature for the two SiPM technologies under study.}
\end{figure}

One of the main contributions to the correlated noise in these SiPM arrays is the presence of direct cross-talk (DiXT). The probability of DiXT was measured against temperature for different $\Delta V$ in the two SiPM array technologies. As the light intensity was set low enough to detect only one photon at a time, setting  DiXT detection threshold at \mbox{$1.17 \times PE$} (pulse amplitude of one photo-electron) and the arriving time after the primary pulse below \mbox{2.0 ns} allows the detection of pulses produced by DiXT. The obtained DiXT probability against temperature for the two types of SiPM arrays under study is shown in figure~\ref{fig:dixt}.

The AP detection threshold was set between \mbox{$[0.6, 0.85] \times PE$} and the arriving time $>15\,ns$. Within the correlated noise, the presence of AP was negligible for both technologies at RT and this remains the same for the Hamamatsu SiPM arrays with cooling. However, a significant increase of AP was observed in the FBK SiPM arrays with cooling. Nevertheless, at 100 K, the contribution of AP and DiXT to the correlated noise are similar.

\section{Summary and conclusions}

A detailed characterization of two different technologies of SiPM arrays, produced by FBK and Hamamatsu, was performed. Key performance parameters were evaluated in a dedicated temperature-controlled cryogenic setup from 300 K down to 100 K. It was found that the $V_{bd}$ decreases linearly with the decreasing temperature until a temperature around 200 K. Below this temperature, the $V_{bd}$ decreases at a lower rate and it shows a quadratic dependence with temperature. Also, in both SiPM array technologies, the quenching resistor increases its value significantly at cryogenic temperatures. Using as a reference its design value at RT ($R_q = 500\,k\Omega$), for the FBK SiPM arrays the value increases by a factor of $\approx5.0$, and for the Hamamatsu ones by a factor of $\approx7.0$. This leads to the same increase in the recovery time of the cell at cryogenic temperatures.

The other parameters studied showed a more moderate variation with cooling. In particular, the PDE remains constant within uncertainties in the full range of temperatures studied. The gain increases between 10\% and 20\% in the FBK SiPM arrays and it does not show a big variation in the Hamamatsu ones, with a measured increase below 10\%. The DiXT probability decreases slightly with cooling in the case of the Hamamatsu SiPM, below 10\%, while the FBK SiPM shows an accentuated decrease of around 20\%.

We can conclude that the performance of the SiPM arrays studied here does not degrade with cooling, showing very similar values of PDE, gain and DiXT probability but some key points can affect the performance operation if they are not properly designed or characterized. First is the correct evaluation of the $V_{bd}$ to operate the SiPM at the optimum $\Delta V$, and second is the proper tuning of the quenching resistor to have the desired recovery times at cryogenic temperatures.

Finally, the presence of AP at cryogenic temperatures in the FBK SiPM arrays is an important point to mention. Although it is not yet fully understood, this is most likely related to the presence of impurities in the material and it represents an important point to take into account during the production process.

\label{sec:summary}


\end{document}